# Unconventional short-range structural fluctuations in cuprate superconductors

D. Pelc[1,2*], R. J. Spieker[1], Z. W. Anderson[1], M. J. Krogstad[3], N. Biniskos[1#], N. G. Bielinski[1+], B. Yu[1], T. Sasagawa[4], L. Chauviere[5,6,7], P. Dosanjh[5,6], R. Liang[5,6], D.A. Bonn[5,6], A. Damascelli[5,6], S. Chi[8], Y. Liu[8], R. Osborn[3], and M. Greven[1*]

[1]School of Physics and Astronomy, University of Minnesota, Minneapolis, MN, USA

[2]Department of Physics, Faculty of Science, University of Zagreb, Bijenička 32, HR-10000 Zagreb, Croatia

[3]Materials Science Division, Argonne National Laboratory, Argonne, IL, USA

[4]Materials and Structure Laboratory, Tokyo Institute of Technology, Kanagawa 226-8503, Japan

[5]Quantum Matter Institute, University of British Columbia, Vancouver, BC V6T 1Z4, Canada

[6]Department of Physics & Astronomy, University of British Columbia, Vancouver, BC V6T 1Z1, Canada

[7]Max Planck Institute for Solid State Research, Heisenbergstraße 1, 70569 Stuttgart, Germany

[8]Neutron Scattering Division, Oak Ridge National Laboratory, Oak Ridge, TN, USA

#present address: Jülich Centre for Neutron Science, Garching, Germany

+present address: Department of Physics, University of Illinois at Urbana-Champaign, IL, USA

*correspondence to: greven@umn.edu, dpelc@phy.hr

**Abstract**

The interplay between structural and electronic degrees of freedom in complex materials is the subject of extensive debate in physics and materials science. Particularly interesting questions pertain to the nature and extent of pre-transitional short-range order in diverse systems ranging from shape-memory alloys to unconventional superconductors, and how this microstructure affects macroscopic properties. Here we use neutron and X-ray diffuse scattering to uncover universal structural fluctuations in $La_{2-x}Sr_xCuO_4$ and $Tl_2Ba_2CuO_{6+\delta}$, two cuprate superconductors with distinct point disorder effects and with optimal superconducting transition temperatures that differ by more than a factor of two. The fluctuations are present in wide doping and temperature ranges, including compositions that maintain high average structural symmetry, and they exhibit unusual, yet simple scaling behavior. The scaling regime is robust and universal, similar to the well-known critical fluctuations close to second-order phase transitions, but with a distinctly different physical origin. We relate this behavior to pre-transitional phenomena in a broad class of systems with structural and magnetic transitions, and propose an explanation based on rare structural fluctuations caused by intrinsic nanoscale inhomogeneity. We also uncover parallels with superconducting fluctuations, which indicates that the underlying inhomogeneity plays an important role in cuprate physics.

**Introduction**

An overarching question in the study of quantum materials is the nature of the relation between structural and electronic properties[1]. This is especially the case for transition metal oxides such as the cuprates, which are lamellar materials prone to structural distortions and complexity[2,3]. Although it was noted early on that purely electronic theories of these high-transition-temperature (high-$T_c$) superconductors are incomplete[2,4], this insight is often overlooked[5,6]. Moreover, there is growing experimental evidence for the ubiquity of inhomogeneity in cuprates and related materials[1-4,7-15], and its consequences for electronic and structural phase transitions have been studied theoretically[16-20]. Yet short-range correlated inhomogeneity is difficult to study experimentally, since most probes yield bulk averages, or do not distinguish between doping-related point defects[21] and more complex nanoscale correlations. Addressing this issue is crucial: recent nonlinear magnetic response measurements point to the importance of inherent structural inhomogeneity for understanding superconducting fluctuations in unconventional oxide superconductors[12,22]; local strains can strongly influence the transition from insulating parent compounds to metallic superconductors[23]; and it is a distinct possibility that electronic excitations within pre-formed structural nanoscale regions mediate the pairing of conduction electrons, and hence drive high-$T_c$ superconductivity[3,24]. The question of short-range structural correlations transcends the field of cuprate physics, and the study of structural fluctuations is essential for understanding a broad range of systems, especially materials that are close to structural instabilities.

Here we study correlated structural inhomogeneity in two cuprates, $La_{2-x}Sr_xCuO_4$ (LSCO) and $Tl_2Ba_2CuO_{6+\delta}$ (Tl2201), by focusing on a specific octahedral tilt distortion. We observe an unusual fluctuation regime, close similarities with superconducting fluctuations, and unexpected links to magnetic fluctuations and martensitic systems. These two cuprate compounds continue be the focus of intense study[11,25-29]. It is well known that LSCO undergoes a transition from high-temperature tetragonal (HTT) to low-temperature orthorhombic (LTO) structure in the underdoped part of the phase diagram[30], and that overdoped Tl2201 displays similar orthorhombic distortions with increasing doping[31]. The distortions involve $CuO_6$ octahedral tilts around the in-plane diagonal, as shown in Fig. 1, with two-fold ground state degeneracy, *i.e.*, two orthogonal tilt orientations are possible. Substantial *local* orthorhombic correlations survive above the structural transition temperature in LSCO[32-34], and are also present in compositions that are nominally tetragonal at all temperatures[31,32]. Tl2201 and LSCO are doped with charge carriers *via* the introduction of interstitial oxygen and the substituton of La with Sr, respectively, which gives rise to qualitatively different point-disorder effects[9,35]. However, we find the same nanoscale orthorhombic fluctuations in the nominally tetragonal phases of both cuprates, including undoped $La_2CuO_4$, with correlation lengths and intensities that follow unusual temperature dependences and doping-independent scaling. Moreover, the universal behavior is strikingly similar to the recently uncovered scaling of superconducting fluctuations in cuprates[13-15] and other oxides[12]. We therefore uncover a generic fluctuation regime that is qualitatively different from the well-known critical fluctuations that lead to second-order phase transitions, but that shows a similar insensitivity to the details of the system. The present findings are relevant to a wide range of materials with structural pretransitional phenomena, such as systems that undergo martensitic transformations[36,37] and strontium titanate[38,39], and they have implications for the general understanding of phase transitions and fluctuations far from the critical point.

We propose that these results signify emergent behavior involving rare spatial-disorder fluctuations, whereby locally orthorhombic patches form already deep in the tetragonal phase due to inhomogeneous interactions. This would imply that an underlying, intrinsic imhomogeneity underpins the unconventional fluctuation regime. Such inhomogeneity has also been suggested to induce an inhomogeneous electronic localization gap that determines much of the normal-state properties of the cuprates[24,40]. While the specific LTO distortion studied here is absent in most cuprates, within the proposed picture it serves as an indirect probe of the underlying inhomogeneity that determines the local interaction landscape.

**Results**

We have performed X-ray and neutron diffuse scattering experiments on high-quality LSCO and Tl2201 single crystals using recently developed, highly sensitive instruments and methodology (for details, see Methods and Supplementary Information). This has allowed us to follow structural fluctuations over a wide range of doping, temperature and energy, and thereby to study their complex dynamical behavior and universal properties hundreds of Kelvin above the tetragonal-orthorhombic transition temperature.

*Static and dynamic fluctuations.* Figure 1 shows raw X-ray diffuse scattering data for LSCO and Tl2201 samples in the overdoped part of the phase diagram. In both cases, the average structure is tetragonal at all temperatures, yet diffuse scattering is observed away from the Bragg peaks. We focus here on the clear diffuse feature at the half-integer LTO superstructure peak positions, which correspond to a $\sqrt{2} \times \sqrt{2}$ unit cell increase. This feature is well known, and a direct signature of LTO tilts. We note that there are two equivalent axes for the LTO tilts, <110> and <1-10> in tetragonal notation, and a locally coherent superposition of tilts around both directions is in principle possible. If the magnitudes of the two orthogonal tilts are equal, the resultant distortion is a tilt around the Cu-O bond direction, known as low-temperature tetragonal (LTT) tilt, and the more general case of unequal magnitudes is referred to as a low-temperature less-orthorhombic (LTLO) tilt[30]. The presence of LTLO distortions would lead to scattering intensity around forbidden *integer* Bragg peak positions of the tetragonal structure. However, additional diffuse scattering of different origin (including phonon scattering and local distortions due to dopants) is present at these positions, and will be discussed in detail in future work.

Diffuse X-ray scattering is an energy-integrating probe, and we performed complementary neutron scattering measurements to gain further insight into the dynamical properties of the LTO correlations. Due to the constraints on sample size and counting times, neutron experiments are not possible with Tl2201. Neutron data were collected for two LSCO samples, on two distinct instruments: a time-of-flight diffuse scattering spectrometer (sample with $x = 0.2$), and a triple-axis spectrometer (sample with $x = 0.155$); see Methods for details. The high-resolution diffuse scattering instrument permits elastic discrimination, *i.e.*, it is possible to extract the quasielastic (below ~0.5 meV) component along with the energy-integrated response (up to ~10 meV). The quasielastic scattering is shown in Fig. 2a for the crystal with $x = 0.2$. It is immediately clear that the diffuse superstructure peaks are split, in contrast to the X-ray data, where a single peak is seen. This stark difference is even more obvious from the diagonal line cuts shown in Fig. 2b, and it indicates a nontrivial evolution with energy of the orthorhombic fluctuations.

The incommensurate peaks were also observed in previous quasielastic triple-axis neutron scattering studies[33,41] and attributed to a spatial modulation of the LTO structure[33]. Here we provide a somewhat different explanation. We model the diffuse scattering from short-range orthorhombic regions as a superposition of static and dynamic components. In order to obtain the quasielastic incommensurate peaks, it is sufficient to assume that the static component has a Gaussian envelope with a single antiphase boundary in the middle (Fig. 2d), where the orthorhombic tilt angle changes sign (see also Methods). No long-range modulation is necessary, or indeed physical, given the short correlation length. In contrast, the dynamic component is taken to have a simple Gaussian tilt-angle profile, with the same characteristic length (Fig. 2c). A linear superposition of the two components then captures the contrast between the neutron and X-ray data well, with the relative weight of the components as the only free parameter (Fig. 2b).

This picture is confirmed and substantiated in more detailed energy-resolved triple-axis neutron measurement of the $x = 0.155$ sample (Fig. 2e-g). Three distinct components are observed: the incommensurate, quasistatic feature also seen in neutron diffuse scattering; a commensurate dynamic feature; and an optical phonon branch with a clear dispersion[42]. Importantly, the intermediate dynamic feature does not simply correspond to the low-energy part of the phonon branch: after normalization with the phonon occupation (Bose) factor, it still shows a significant temperature dependence, whereas the signal above about 5 meV does not, as expected for phonons (Fig. 2h). This implies that a large fraction of the spectral weight below 5 meV originates from local modes around the LTO position; as shown in Fig. 2h, the signal in the 1 to 5 meV range approximately follows the temperature dependence of the integrated quasielastic peak weight. We note that similar local modes were found for LTLO-type distortions in LSCO with $x = 0.07$ below the HTT-LTO transition[43].

We emphasize that our simple calculation uses a minimal number of assumptions. The correlation length is assumed to be the same for the static and dynamic component, while it is clear from Fig. 2e-g that the dynamic feature is narrower than the quasielastic feature (by a factor ~3 along the [H H 2] direction). Moreover, the actual tilt patterns might be more complicated. It has been suggested from pair distribution function experiments on powder samples that LTLO tilts are always present locally, at least at low temperatures[44]. In our single crystal data the LTO and LTLO contributions are clearly separated in reciprocal space, and a weak signal also appears, *e.g.*, at the forbidden Bragg position position (3 0 4), as shown in Fig. 2a. The integrated intensity is much smaller than the LTO feature, implying that, far above the structural transition temperature, the quasistatic tilts are predominantly LTO. As noted, additional diffuse scattering from other distortions is possible at the forbidden Bragg positions. However, the weights at half-integer and integer positions are comparable in the total scattering channel (see Supplementary Figure S4), which raises the interesting possibility that the dynamic fluctuations involve a sizable LTLO component. Temperature-dependent energy-resolved measurements similar to those shown in Fig. 2e-h would be necessary to distinguish between local modes and phonon scattering. Yet such an analysis for a sample with lower doping[43] suggests the local LTLO-like fluctuations decay significantly already below the HTT-LTO transition temperature, and are thus likely not relevant far above the transition.

Two important conclusions can be drawn from our neutron results. First, the LTO correlations are not purely static, but present in a fairly wide energy range, with an interesting energy dependence. In particular, the static configuration with an antiphase boundary likely lowers the overall elastic energy of the system, given that it is embedded in a tetragonal matrix, while the fluctuations at nonzero energy are not as constrained. Most importantly, both components represent aspects of the same phenomenon, as evidenced by their similar temperature dependences (Fig. 2h). The energy-integrated X-ray data are thus a robust measure of these local correlations, which will be important for our analysis of the temperature and doping dependences below. Second, an incommensurate response can be obtained without a well-defined modulation wavevector, and the incommensurability is simply determined by the underlying correlation length. This observation may be relevant for a wide range of systems with short-range correlations. Interestingly, the X-ray diffuse peaks are slightly shifted from the nominal commensurate position (see Supplementary Information), which might indicate a local expansion of the lattice within the orthorhombic regions. The shifts are easier to discern in higher Brillouin zones and not observed in neutron scattering, likely due to the smaller reciprocal lattice vectors involved.

*Universal behaviour.* Diffuse scattering around the superstructure positions is found in LSCO at all studied doping levels and extends to high temperatures (Fig. 3 and Supplementary Figure S5). Such a broad temperature range for structural fluctuations is unusual, even in materials with a martensitic transition, where pre-transitional effects have been extensively documented[37]. Furthermore, the temperature dependence of the integrated LTO diffuse intensity is consistent with an exponential decay, and the decay rate of about 1/(200 K) is nearly universal across the LSCO phase diagram and for Tl2201 above the superconducting $T_c$. The rate seems to show a small systematic increase with doping for LSCO (Supplementary Figure S5), which may be related to disorder introduced by Sr substitution, or changes in the ratio of static and dynamic contributions. Interestingly, the structural data for Tl2201 show a change around $T_c$. We were unable to test this in overdoped LSCO, since the $T_c$ values are below the experimental base temperature of 30 K. For LSCO, the widths of the diffuse peaks scale on a master curve (Fig. 3b), consistent with $(T - T_{LTO})^{1/3}$. The Tl2201 peak widths are also consistent with the same curve if we assume that the effective $T_{LTO}$ is roughly $T_c$, *i.e.*, that the system would undergo a structural phase transition around $T_c$ if it were not for the appearance of bulk superconductivity. Yet there is no direct experimental support for this assumption, and it is possible that the intrinsic size of the correlated regions is somewhat different for the two cuprates. The Tl2201 correlation lengths would then have to be multiplied by a constant factor to be comparable to LSCO. Since this factor is *a priori* unknown, little can be said about the effective $T_{LTO}$ for the Tl2201 sample.

Similar to the extracted correlation lengths, the normalized diffuse intensities collapse on the same curve upon shifting the temperature by $T_{LTO}$ (Fig. 4 and Supplementary Information). The correlation length and intensity scaling implies that the effect of doping is predominantly a simple shift of $T_{LTO}$; the short-range correlations are otherwise insensitive to the Sr concentration. We note that previous cold neutron results are consistent with an approximately doping-independent exponential dependence of the diffuse superstructure peak intensities[33]. Our measurements of the temperature dependence of the elastic neutron signal for $x = 0.155$ agree very well with the X-ray results (inset in Fig. 3a). Together with the energy-resolved measurements discussed above, this shows that it is justified to take the energy-integrated response from X-ray scattering to be a good measure of the orthorhombic fluctuation intensity. Moreover, the slope of the exponential decay is independent of the X-ray or neutron wavevector, which implies that the influence of temperature-dependent Debye-Waller factors is negligible (see also Supplementary Information and methods for more details).

The observed behavior is simple, but unusual: neither exponential decay nor scaling with relative temperature ($T - T_{LTO}$) is expected for fluctuations leading to a second-order phase transition, which would be associated with power-law dependences in the reduced temperature ($T - T_{LTO})/T_{LTO}$. We emphasize that the exponential fluctuation regime exhibits scaling properties and insensitivity to system details similar to critical fluctuations, but due to the extended temperature range and exponential behavior, its physical origins must be qualitatively different. We explore here a possible explanation through rare large spatial disorder fluctuations (rare-region effects), which would naturally lead to an exponential-like dependence. Rare-region effects beyond mean-field theory have long been discussed in the context of magnetism[45-47]. A related phenomenon is the optical Urbach effect, which results from exponential tails in the density of states of semiconductors and is observed in a wide range of materials[48,49]. The physical picture is as follows. At the mean-field level, the LTO transition at $T_{LTO}$ occurs due to a lattice instability, which originates in the specific body-centered layered perovskite structure of LSCO and Tl2201. This instability can be expressed through an effective coupling strength within a Ginzburg-Landau-Wilson Hamiltonian. However, if some underlying inhomogeneity induces spatial variations of the effective coupling, exponentially rare locally-ordered regions will form above $T_{LTO}$, since the local coupling can be stronger than the average coupling that causes the bulk phase transition. In addition, given the double degeneracy of the LTO order parameter, inhomogeneity can act as a random field on the Ising-like tilt variable[20]. Conventionally, quenched point disorder is taken to generate the local coupling/field landscape that leads to rare-region fluctuations[17,19,20,46], but in the present case it seems unlikely that disorder induced by chemical doping plays that role: the observed exponential behavior is nearly doping independent in LSCO, and appears even in undoped, nominally stoichiometric $La_2CuO_4$. Furthermore, comparable behavior is observed for Tl2201, which exhibits qualitatively different point disorder effects. Instead, the required spatial variation could be generated by some intrinsic, hidden inhomogeneity, which is therefore indirectly revealed by our experiments. Such inhomogeneity has also been argued to lead to a common superconducting fluctuation regime in oxide superconductors, including the cuprates[12-14]. However, as discussed below, local electronic correlations likely play a prominent role as well.

For classical, thermal phase transitions, the thermodynamic signatures of rare-region effects are difficult to observe[46]. From our experiments, it is not possible to directly determine if the diffuse LTO superstructure peaks indeed originate from strong and rare fluctuating regions, or if the fluctuations are approximately homogeneous in real space. Yet it is possible to perform a consistency check based on the simple modeling described above. From this, we estimate that the typical distance between ordered regions is about 60 unit cells far above $T_{LTO}$ (see Supplementary Information for details), where the typical correlation length is 4-5 unit cells, *i.e.*, the regions are indeed far apart. The dependence of the density of rare regions on the "distance" (in temperature) from the bulk phase transition is exponential to leading order within a generic Landau-Ginzburg-Wilson theory of thermal phase transitions with a Gaussian distribution of the local coupling strengths[46]. The exponential decay rate is then a measure of the distribution width. Further, there exists a characteristic length scale, $L_{min}$, that corresponds to the minimum ordered-region size, and that appears because of finite-size scaling; fluctuations prevent order in regions smaller than $L_{min}$. The scattering then is dominated by regions of size $\sim L_{min}$, since the probability of finding larger regions quickly decreases with their size. Note that $1/L_{min} \sim r_0^\nu$, where $r_0 \sim T - T_{LTO}$ is the distance from the bulk phase transition, and $\nu$ is the correlation length exponent. If $r_0$ lies outside the narrow critical region, it is appropriate to use the mean-field value[46] $\nu = 1/2$, which implies a square-root dependence of the diffuse peak width on relative temperature. Indeed, we find that the widths scale with a power of $T - T_{LTO}$, without normalization of the temperature scale by $T_{LTO}$. This is expected within the rare-region scenario if an underlying inhomogeneity sets the distribution width independently of $T_{LTO}$. The exponent appears to be smaller than 1/2, however, and consistent with 1/3 (Fig. 3b). This value is not easy to understand, since to our knowledge no universality class has $\nu < 1/2$. This disagreement could indicate that $r_0$ is a more complicated function of the relative temperature; or that one should take a *lower* value $T_{LTO}$' as the reference temperature; or that the underlying physical picture is different. Since there are indications from NMR[50] and neutron scattering[51] experiments that the LTO transition is weakly first-order in doped LSCO, it is plausible that the transition occurs at a nonzero value of $r_0$ in the Landau-Ginzburg-Wilson theory, leading to a shifted $T_{LTO}$'. In fact, a weakly first-order transition has been discussed in the context of martensitic materials and captured theoretically upon inclusion of terms in the free energy that couple the order parameter to strain[37]. As seen from Fig. 3b, $T_{LTO}$' $\sim T_{LTO} - 25$ K with $\nu = 1/2$ provides an approximate description of the data. The rare-region picture provides a natural way to reconcile the different microscopic insights into the LTO transition gained from NMR[50] and X-ray[32] experiments for LSCO. The transition is clearly predominantly displacive, in agreement with [139]La NMR, but with the vestigial correlations above $T_{LTO}$ reminiscent of an order-disorder scenario, as suggested by pair-distribution-function analysis[3,32]. However, a more quantitative understanding that includes random field effects is beyond the scope of this work.

**Discussion**

We first discuss some immediate implications of the finding of LTO fluctuations and their scaling behaviour to recent work on LSCO and Tl2201. The prevalence of short-range LTO regions throughout the phase diagram suggests that they might also be present in nominally tetragonal LSCO films and heterostructures. This could, in turn, affect electronic properties such as transport anisotropy[26], especially if one or more sample dimensions approach $L_{min}$. The corresponding superstructure peaks might be detectable in dedicated experiments on films. The low-temperature lattice specific heat of bulk LSCO exhibits a peak around the doping level where $T_{LTO}$ goes to zero[52], but to our knowledge, this has not been investigated in detail. In principle, pre-transitional effects could influence the low-temperature specific heat[46] and complicate the separation of lattice and electronic contributions[27]. The low-temperature regime, where quantum structural fluctuations become important, could be interesting in its own right. The change in LTO diffuse intensity below $T_c$ in Tl2201 (Fig. 3) demonstrates the possibility of a coupling of LTO correlations to superconductivity. Comparable conclusions were obtained for in-plane buckling distortions from X-ray absorption fine structure and powder pair distribution function measurements[3,53]. Moreover, it was recently suggested that disorder influences the disappearance of superconductivity at high doping[24,54], with an emergent percolative regime. For LSCO, this occurs below $x \sim 0.26$. Intriguingly, at that doping level, the low-temperature orthorhombic correlation length is about 2 nm, similar to the superconducting coherence length, another indication of an emergent relation between the LTO and superconducting correlations. We also note that short-range structural fluctuations could also be present at much higher energies than accessible in the present study, with possible relevance for superconducting pairing.

A local structural symmetry that is lower than the average symmetry will have important qualitative repercussions for the electronic subsystem. Such effects have been shown to be essential for the physics of manganites and other oxides[1,23], and they have been discussed in the context of cuprate pseudogap formation and superconducting pairing[3,4]. Another case in point is electronic nematicity – the spontaneous lowering of the electronic symmetry – which has been extensively studied in iron-based and cuprate superconductors[16,17,19,26,55]. It is a distinct possibility that structural precursor effects play a role in cuprate electronic symmetry breaking, even in compounds such as simple tetragonal[56] $HgBa_2CuO_{4+\delta}$. A close relation between the orthorhombic distortions and electronic features such as spin and charge stripes in lanthanum-based cuprates was proposed in previous work[33,44,57], with a notable similarity between the LTO and spin-stripe incommensurability values[33]. Our investigation focuses on structural degrees of freedom and cannot provide direct evidence either for or against causal relations to electronic ordering tendencies. Yet our modeling of the LTO incommensurability suggests that this can be explained simply as an antiphase boundary effect, which would imply that the spin stripe similarity is probably accidental, although it is possible that electronic degrees of freedom enter the overall energy balance that leads to the formation of the boundary[57]. We note, however, that spin-stripe order in La-based cuprates melts at significantly lower temperatures than $T_{LTO}$ in underdoped compounds, is very weak or absent beyond optimal doping, and absent in the undoped system.

More broadly, rare-region effects have been most extensively discussed in the context of magnetic systems and, more recently, superconductors[12,54,58]; they might also occur in ferroelectrics, despite the presence of long-range dipolar interactions[59]. The extension to structural phase transitions that we suggest here could be of wide relevance. Prominent examples include the long-standing question of the 'central peak' and pre-transitional static order above the cubic-tetragonal transition in strontium titanate[38,60] as well as short-range correlations and patterns in systems with martensitic transitions. While mean-field theories of martensitic precursors are well developed[36,37], our results indicate that rare fluctuations could play an important role, especially far above the transition. We see no direct evidence of mesoscopic inhomogeneity such as tweed patterns, which would appear as structured diffuse scattering around the Bragg positions, and which have in fact been observed in related, intentionally-disordered systems[2,36] such as $YBa_2(Cu_{1-x}Al_x)_3O_{7-\delta}$. The appearance of antiphase boundaries within the distorted regions could be viewed as incipient mesoscopic correlations. We note that the anisotropic random-field Ising model[20] might provide an effective microscopic framework to understand our results: depending on the disorder strength and in-plane/out-of-plane coupling anisotropy, this model can show an extended regime with locally ordered regions, and the physics is insensitive to the details of the random field, in line with our observations. The model has been used, *e.g.*, to study the effects of quenched disorder on charge stripe[20] and Ising-nematic order[16,17] in oxides, as well as structural fluctuations in FeSe-based compounds[61], which show qualitatively similar features to the LTO fluctuations observed here. Yet we are not aware of calculations or theoretical predictions that could be directly compared to our measurements well above the ordering temperature, in particular the peculiar scaling relations that we have uncovered. More detailed theoretical work is needed to explicitly study the high-temperature fluctuations.

Regardless of the validity of the rare-region explanation, the exponential intensity decay and scaling are striking. Yet perhaps the most interesting result, shown in Fig. 4, is the close correspondence with the doping- and compound-independent exponential decay of superconducting fluctuations in different cuprates[12-15] (including LSCO and simple-tetragonal $HgBa_2CuO_{4+\delta}$). Since the LTO distortion is only present in some cuprates (*e.g.*, it is absent in $HgBa_2CuO_{4+\delta}$), the LTO correlations are certainly not the underlying cause for the universal exponential superconducting fluctuation behaviour. Instead, the closely similar temperature dependences suggest a generic phenomenon, similar to the universal features of critical fluctuations close to a second-order phase transition. In the proposed rare-region scenario, the same inhomogeneity is the cause of both LTO and superconducting fluctuation behaviour.

The LTO fluctuations are effectively an indirect probe of the underlying energy landscape, and therefore do not provide direct insight into its nature. However, the simple observed scaling along with insights from previous work allow some informed speculation. As noted, in a rare-region picture, the scaling in Fig. 4 would indicate that underlying inhomogeneity is similar in different cuprate families and roughly independent of doping, and thus at best weakly related to doping-induced point disorder. This is emphasized by the fact that stoichiometric $La_2CuO_4$ shows the same exponential decay and scaling as the doped compounds, which is not easy to understand in any point-defect-based model. However, all cuprates share the $CuO_2$ layers as a common structural element and are prone to distortions due to their perovskite-based structure. The atomic size differences between adjacent perovskite and rock-salt layers induce internal strain, and Jahn-Teller instabilities can also cause distortions of the $CuO_6$ octahedra[1,3,4]. It would therefore be unsurprising to find intrinsic inhomogeneity that is rooted in the structural properties of the cuprates. There is a distinct possibility that the cuprates (and related compounds) are never in thermodynamic equilibrium[62], with a wealth of metastable states that are close in energy. Dynamic inhomogeneity would then appear at high temperatures due to thermal activation and become arrested upon cooling; a similar mechanism has been evoked to explain exponential Urbach tails in amorphous semiconductors[63]. We also note that micro-X-ray experiments revealed fractal disordered structures[10] in $La_2CuO_{4+\delta}$ and that fractal patterns have also long been discussed in systems with martensitic transitions[64] and for ordered domains in the random-field Ising model[65]. It is therefore possible that scale-free structural inhomogeneity is present in the cuprates and other perovskite-based materials. Yet, in the cuprates, the coupling between structure and the electronic system might be particularly important, through a complex interplay between long-range elastic interactions and more localized electronic correlations[57]. We speculate that the universal aspect of the inhomogeneity arises *via* a self-organized process that involves both the electronic and atomic subsystems, and that it could play a pivotal role in the physics of the normal and superconducting states. Indeed, evidence is accumulating for the existence of distinct Fermi-liquid-like and local pseudogap-related, disordered components of the electronic subsystem[24,28,40,66,67]. This was quantified in a recent phenomenological model through a nearly compound-independent inhomogeneity of hole-localization gaps, that leads to the gradual crossover from a large Fermi surface at high doping and/or high temperature to Fermi arcs in the pseudogap state[24,40]. It is possible that self-organized hidden inhomogeneity creates an interaction landscape that leads to the two-component behavior and universal localization-gap inhomogeneity. Other emergent phenomena, such as LTO fluctuations, superconducting fluctuations, intra-unit-cell magnetism, electronic nematicity, and short-range spin, charge and pair density waves[6], would then also develop within this landscape. Targeted experiments are needed to directly observe the hidden inhomogeneity, and its microscopic understanding may hold the key to explain the unusual properties of the cuprates.

**Methods**

*Samples.* Single crystals of $La_2CuO_4$ (LCO; $x = 0$) and LSCO were grown with the traveling-solvent floating-zone technique. The doped LSCO crystals were annealed post-growth at 800°C in air for 24-48 hours, characterized with SQUID magnetometry, and exhibit $T_c$ values consistent with the nominal Sr concentrations[68]. The Néel and structural transition temperatures of LCO are known to be very sensitive to oxygen off-stoichiometry. Our LCO crystal was reduced at 900°C in vacuum for 30 minutes to remove excess oxygen[69]. A subsequent SQUID magnetometry measurement of the antiferromagnetic transition yielded $T_N \sim 325$ K, and our X-ray scattering data revealed $T_{LTO} \sim 530$ K. These high transition temperatures indicate that our LCO sample was stoichiometric, *i.e.,* undoped[69]. The high quality of the crystals was further confirmed in the diffuse scattering measurements, which showed sharp Bragg peaks with no visible mosaicity (with the exception of the ~800 mg crystal with $x = 0.2$ used for diffuse neutron scattering, where slight sub-grain misorientation was observed − Fig. 2 and Supplementary Figure 3). Samples for X-ray scattering had masses in the 1-10 mg range, with the Bragg peak widths typically limited by the resolution of the detector. The $x = 0.155$ sample for triple-axis neutron spectroscopy was composed of 4 co-mounted single-grain pieces with total mass of approximately 7.5 g and a Bragg-peak mosaic spread of less than 0.6°. Single crystals of Tl2201 were grown and annealed using established procedures[31], and the sample used for the experiments had a sharp superconducting transition and no visible mosaic.

*Diffuse neutron and X-ray scattering.* Hard X-ray diffuse scattering was measured over large volumes of reciprocal space at beamline 6-ID-D of the Advanced Photon Source, Argonne National Laboratory, using the methods described in detail in [70]. The high photon flux and detector sensitivity lead to short acquisition times, which enabled us to efficiently study one

Tl2201 and a number of LSCO crystals with different Sr concentration levels in wide temperature ranges. Neutron diffuse scattering was performed with the CORELLI instrument[71] at the Spallation Neutron Source, Oak Ridge National Laboratory for LSCO ($x = 0.2$) at two temperatures, 10 K and 300 K. The sample has an estimated LTO transition temperature of 80 K, so the only temperature relevant for short-range LTO correlations is 300 K. The unique design of the CORELLI cross-correlation spectrometer enabled discrimination between quasi-elastic and dynamic scattering. The Mantid software package was used for data reduction, including Lorentz and spectral corrections[72]. The LTO diffuse signal was highest in the $L = 4$ r.l.u. plane (shown in Fig. S3) and barely discernible or absent in other planes, so the $L = 4$ r.l.u. plane was the only one used for analysis.

The superstructure peak intensities in Fig. 3a were obtained by integrating a cubic reciprocal space volume of width 0.3 r.l.u. in all three principal directions and centered on the LTO superstructure position. The thermal and point disorder scattering background was obtained by using a similar integration box in a reciprocal space volume adjacent to the LTO integration box. The LTO positions used to obtain Fig. 3 were (-9 -6 1) for 8% Sr, (-8 -9 1) for 12% Sr (orthorhombic notation); (1.5 7.5 2) for 20% Sr, (4.5 7.5 9) for 24% Sr, (-4.5 7.5 9) for 27% Sr and (-3.5 -3.5 2) for LCO (tetragonal notation); and (2.5 -7.5 9) for Tl2201 (tetragonal notation). The intensity of the diffuse scattering in principle includes a Debye-Waller factor similar to conventional Bragg scattering, that originates from random thermal atomic displacements and has the approximate form $\exp(-q^2 u^2)$, where $q$ is the wavevector and $u$ the average thermal displacement. However, in our case the contribution of the Debye-Waller factor to the temperature dependence is negligible, as evidenced by the fact that the slope of the exponential decay is similar in different Brillouin zones (different values of $q$). This is demonstrated in Fig. S6 for the sample with 24% Sr, where the standard deviations can also be determined in a systematic manner. We note though that the main source of uncertainty for the intensities is not random noise, but systematic errors that originate in setup imperfections – e.g. detector artefacts/streaking and slight sample movement, especially at high temperatures. Several of these artefacts are visible in Fig. 1b,c, including different forms of streaking and dead pixels.

The intensities in Fig. 3a are normalized by the integrated intensities of the nearest allowed Bragg reflections at 30 K to compensate for different sample volumes and X-ray absorption. Due to the relatively limited reciprocal space resolution of 0.02 r.l.u. and the possibility of nonlinearities due to detector overload, the normalization procedure has a systematic error that is difficult to estimate. The Gaussian widths in Fig. 3b are obtained by fitting the sum of a Gaussian peak and third-order polynomial background to the LTO peaks integrated along $H$. The correlation length is calculated as $L = 2.35 / \sigma$, where $\sigma$ is the Gaussian width (standard deviation), and the numerical factor is obtained from the calculation of diffraction from a slab with linear size $L$.

*Neutron triple-axis spectroscopy.* The experiments were performed at the HB-3 instrument of the High-Flux Isotope Reactor (HFIR) at Oak Ridge National Laboratory, with collimations 48'-40'-40'-120'. The data shown in Fig. 2 were obtained in energy-loss mode using a fixed final energy of 14.7 meV and energy-transfer values 0, 1.25, 2.5, 3.75, 5, 7.5 and 10 meV, with scans in constant-energy mode in the [$H\ H\ L$] scattering plane and counting times of up to 4 minutes per point. The FWHM energy resolution was estimated to be 1.3 meV at the elastic line, calibrated with a vanadium standard. Figs. 2e-g were plotted using standard Matlab interpolation, with the acoustic phonon dispersions of the form

$$\omega_q^2 = \omega_{LTO}^2 + Aq^2$$

where $q$ is the wavevector along [HHL] relative to the half-integer point, $\omega_{LTO}$ is the zone-boundary optical phonon frequency, and $A$ a constant. The phonon energies $\omega_{LTO}$ were estimated from published values[42] for a sample with $x = 0.14$ to be 4.5, 5 and 6 meV for 250, 300 and 400 K, respectively, and $A = 8000$ for all temperatures. The LTO transition temperature was estimated to be 180 K. In order to obtain the dynamic susceptibility $\chi''$, the datasets at each nonzero energy transfer were normalized by the phonon occupation factor

$$n_s + 1 = \frac{1}{1 - e^{-\hbar\omega/kT}}$$

where $\hbar\omega$ is the energy transfer, $k$ the Boltzmann constant and $T$ the temperature. The procedure is similar to previous work, *e.g.* on tellurides[73]. The two sets of results in Fig. 2h (full and empty circles) were then obtained by a double summation of the data along [H H 2] and energy in the specified ranges, at each temperature. The elastic line intensity (squares) was calculated as sums of the zero energy-transfer datasets along [H H 2], with no normalization. The reciprocal space summation range was 1.5±0.2 r.l.u. for all datasets.

*Modeling of LTO diffuse peaks.* The diffuse scattering calculation shown in Fig. 2b is performed numerically as a direct lattice scattering sum, using orthorhombic coordinates and a simple parametrization of the LTO distortion, as follows. The primitive $La_2CuO_4$ unit contains one Cu atom, two in-plane and two apical oxygen atoms, and two lanthanum atoms (we do not take La/Sr substitution into account). Each unit cell in a square two-dimensional lattice is ascribed a local LTO tilt angle α, with relative atomic coordinates in each octahedron parametrized as

$$O_{11} \to \begin{pmatrix} 0 \\ -0.184c/a\sqrt{2}\sin\alpha \\ 0.184\cos\alpha \end{pmatrix}; O_{12} \to \begin{pmatrix} 0 \\ 0.184c/a\sqrt{2}\sin\alpha \\ -0.184\cos\alpha \end{pmatrix},$$

$$O_{21} \to \begin{pmatrix} 1/4 \\ -1/4 \\ -a/2c\sin\alpha \end{pmatrix}; O_{22} \to \begin{pmatrix} 1/4 \\ 1/4 \\ a/2c\sin\alpha \end{pmatrix},$$

$$La_{11} \to \begin{pmatrix} 0 \\ 0.361c/a\sqrt{2}\sin\gamma\alpha \\ 0.361\cos\gamma\alpha \end{pmatrix}; La_{12} \to \begin{pmatrix} 0 \\ -0.361c/a\sqrt{2}\sin\gamma\alpha \\ -0.361\cos\gamma\alpha \end{pmatrix},$$

where the apical and in-plane oxygen atoms are referred to as $O_{1(1,2)}$ and $O_{2(1,2)}$, respectively, and the Cu atom is taken to be at the origin; $a = 3.78$ Å and $c = 13.2$ Å are the tetragonal unit cell lattice parameters. The factor $\gamma$ takes into account the fact that the heavy La atoms move significantly less than the apical oxygens, with $\gamma \sim 0.1$ taken from crystallographic data on the LTO phase[74]. Each orthorhombic unit cell contains four $La_2CuO_4$ units, and the local tilt angle is taken to be the same for all units within one cell.

The scattering intensity is calculated from a single LTO domain using a direct lattice sum of the form

$$I = \left| \sum_{j,l} f_{j,l}(|\boldsymbol{Q} + \boldsymbol{q}|) e^{2\pi i (\boldsymbol{Q}+\boldsymbol{q}) R_j} \right|^2$$

where the sum goes over all lattice sites $j$, the index $l$ denotes the atom type (Cu, O or La), and the wavevector is decomposed into the reciprocal lattice unit vector $\boldsymbol{Q}$ and the relative wavevector $\boldsymbol{q}$. The wavevector dependences of the atomic form factors $f$ are taken from [75]. The calculation presupposes that kinetic diffration theory is applicable, which is a good assumption for short-range correlations, but possibly not for Bragg scattering. Periodic boundary conditions are assumed for the summation to eliminate finite-size artefacts.

To obtain short-range LTO correlations, we use a local tilt angle distribution that is a sum of two contributions: dynamic (with a simple Gaussian spatial dependence) and static (with an antiphase boundary, i.e. the tilt changing sign):

$$\alpha(x, y) \sim \alpha_0 \exp[-(x^2 + y^2)/\delta^2]\,[\eta + \mathrm{erf}(x)]$$

where $\alpha_0$ is the tilt angle amplitude, $x$ and $y$ are real-space coordinates in relative (orthorhombic) units, erf denotes the Gaussian integral (error function) and represents the static contribution with an antiphase boundary at $x = 0$, and $\eta$ is the relative amplitude of the dynamic (pure Gaussian) contribution. The results shown in Fig. 2 are obtained on a 100x100 two-dimensional grid, with $\delta = 4$ orthorhombic l.u. This value is roughly 1.4 times larger than the correlation length obtained from simple Gaussian fits to the LTO line profiles in reciprocal space, shown in Fig. 3b. The two lengths are defined slightly differently: $\delta$ is the real-space Gaussian width of the LTO regions, whereas the correlation length $L$ is calculated under the assumption that the LTO regions have a boxcar-like profile in real space. The numerical agreement between the two is therefore satisfactory. The results do not substantially depend on the value of $\alpha_0$, so we have used the typical[44] long-range LTO tilt of 5°. Smaller values of $\alpha_0$ essentially only change the intensity of the diffuse signal.

**Acknowledgements**

We thank D. Robinson and S. Rosenkranz for assistance with X-ray scattering experiments, and A. Klein, R. M. Fernandes, C. Leighton, B. Chakoumakos, N. Barišić, R. D. James and B. Shklovskii for discussions and comments. The work at the University of Minnesota was funded by the U.S. Department of Energy through the University of Minnesota Center for Quantum Materials, under Grant No. DE-SC0016371. The work at Argonne was supported by the U.S. Department of Energy, Office of Science, Basic Energy Sciences, Materials Sciences and Engineering Division. The work in Zagreb was supported by Croatian Science Foundation Grant No. UIP-2020-02-9494. A portion of this research used resources at the High Flux Isotope Reactor and Spallation Neutron Source, DOE Office of Science User Facilities operated by the Oak Ridge

National Laboratory. This research used resources of the Advanced Photon Source, a U.S. Department of Energy (DOE) Office of Science User Facility operated for the DOE Office of Science by Argonne National Laboratory under Contract No. DE-AC02-06CH11357. Work at the University of British Columbia was undertaken thanks in part to funding from the Max Planck-UBC-UTokyo Centre for Quantum Materials and the Canada First Research Excellence Fund, Quantum Materials and Future Technologies Program, in addition to the Natural Sciences and Engineering Research Council of Canada's (NSERC's) Steacie Memorial Fellowships (A.D.); the Alexander von Humboldt Fellowship (A.D.); the Canada Research Chairs Program (A.D.); NSERC, Canada Foundation for Innovation (CFI); British Columbia Knowledge Development Fund (BCKDF); and the CIFAR Quantum Materials Program.

**Author contributions**

DP and MG conceived the research; BY, DP, MJK, NB, RO and ZWA performed the X-ray scattering experiments and analyzed data; DP, MJK, NB, SC, RS, YH and ZWA performed the neutron scattering experiments and analyzed data; NGB, RS and TS grew and characterized LSCO single crystals; LC, PD, RL, DAB and AD grew and characterized the Tl2201 crystals; DP and MG wrote the paper, with input from all authors.

**Competing Interests**

The Authors declare no Competing Financial or Non-Financial Interest.

**Data Availability**

All data, materials and computer code used to generate the results in the paper are available from the corresponding authors upon request.

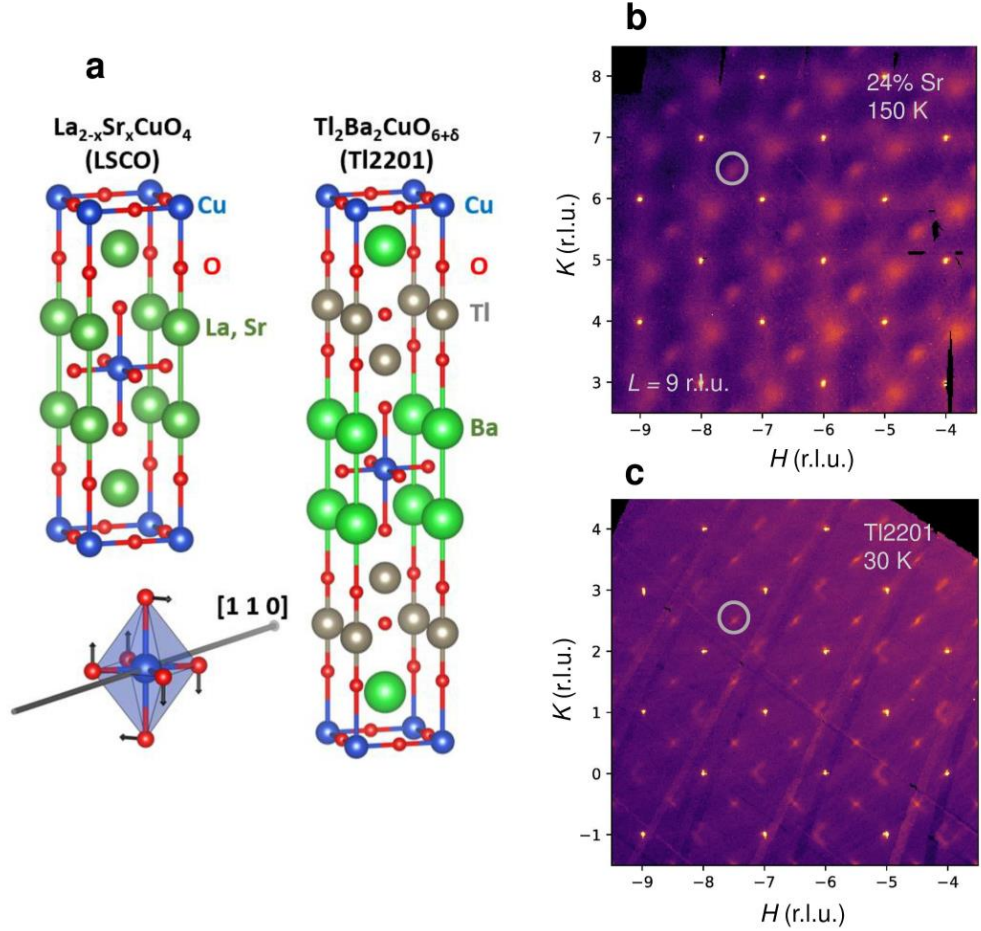

**Fig. 1: Short-range orthorhombic correlations in LSCO and Tl2201. a** Crystal structures of LSCO and Tl2201. Both compounds crystallize in a body-centered tetragonal structure at high temperatures. The LTO distortion corresponds to tilts of the $CuO_6$ octahedra, leading to a doubling of the unit cell volume. **b,c** Diffuse X-ray scattering for (**b**) LSCO ($x = 0.24$) and (**c**) slightly overdoped Tl2201. In both cases, the average structure is tetragonal at all temperatures. The $HK9$ planes are shown, and clear diffuse peaks are observed at reciprocal space positions corresponding to the LTO superlattice (white circles). Black regions are detector artefacts that have been masked. Additional diffuse scattering of a different origin is present, including around forbidden integer Bragg positions. The sharp Bragg peaks correspond to the average tetragonal structure and demonstrate the high quality of the crystals. The scattering in the LSCO sample is shown at 150 K, well above the superconducting $T_c = 15$ K. For Tl2201 we show data at the experimental base temperature of 30 K, far below $T_c = 89$ K.

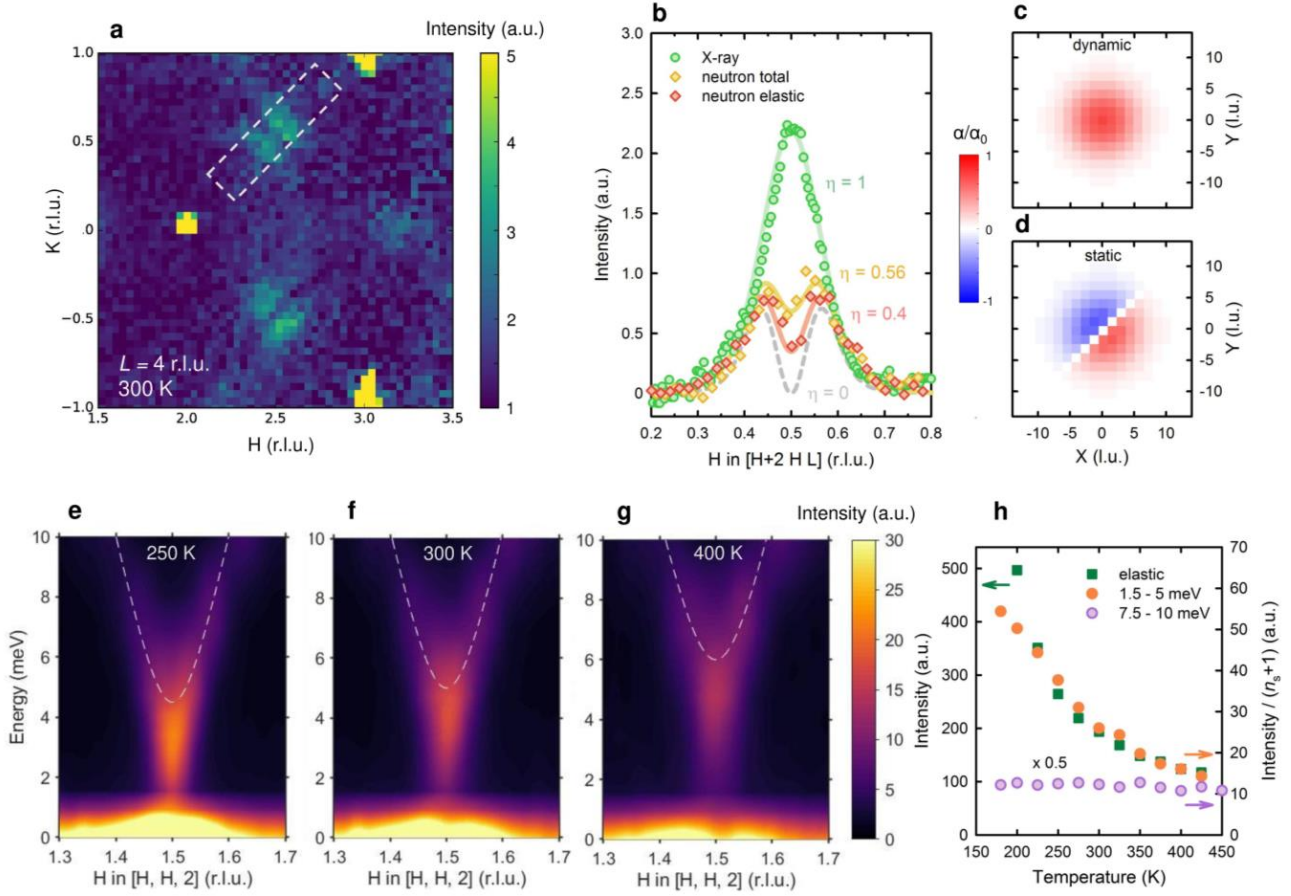

**Fig. 2: Static and dynamic correlations in overdoped LSCO. a** Quasielastic diffuse neutron scattering in a $x = 0.20$ crystal at 300 K, far above the HTT-LTO structural transition of about 60 K. The $HK4$ plane is shown, integrated within a window of $L = 4 \pm 0.2$ r.l.u. (for an expanded view, see Fig. S3). LTO incommensurate diffuse peaks centered at $(2.5, \pm 0.5, 4)$ are clearly visible, along with a weak signal around the forbidden Bragg position $(3\ 0\ 4)$ that likely corresponds to LTLO tilts. **b** Integrated intensity through the superstructure peak along $[H\ 2+H\ 4]$ (marked with the box in (**a**)), for total and quasielastic neutron diffuse scattering (diamonds), as well as X-ray scattering at a similar doping level (circles). The total and quasielastic neutron data are normalized to the respective intensities of nearby Bragg peaks, whereas the X-ray data are scaled to match the low-$H$ tail of the neutron peaks. Polynomial backgrounds have been subtracted from all three datasets. Lines are calculated diffuse scattering from a single nanoscale orthorhombic domain with local orthorhombic tilt angle $\alpha$, and a superposition of a static contribution with an antiphase boundary (shown in real space in **d**) and a dynamic Gaussian contribution (shown in **c**). The chosen maximum value $\alpha_0 = 5°$ corresponds to the typical low-temperature long-range tilt[34,44], the constant $\eta$ is the relative amplitude of the dynamic contribution, and the same orthorhombic correlation length of about 6 tetragonal unit cells is used for all four curves. The in-plane tetragonal lattice unit is 3.78 Å. **e-g** Energy-resolved triple-axis neutron scattering data across the $(1.5\ 1.5\ 2)$ LTO diffuse peak, for a $x = 0.155$ sample ($T_{LTO} = 180$ K), measured at 250 K, 300 K and 400 K,

respectively. The lowest-lying optical phonon branch (dashed lines, estimated from ref. 42 – see Methods) exhibits a local minimum and is seen to disperse upward in energy. Additional spectral weight is present below the phonon branch. **h** Energy- and $H$-integrated intensity for the data in (f-g) corrected by the phonon Bose factor $n_s + 1$ (see Methods), showing excess spectral weight below about 5 meV (full circles). This is evidence for a signal from local fluctuations that likely has the same origin as the $H$-integrated quasielastic incommensurate peak (squares). On the other hand, the weight above 5 meV is temperature-independent (purple circles), as expected for a phonon branch.

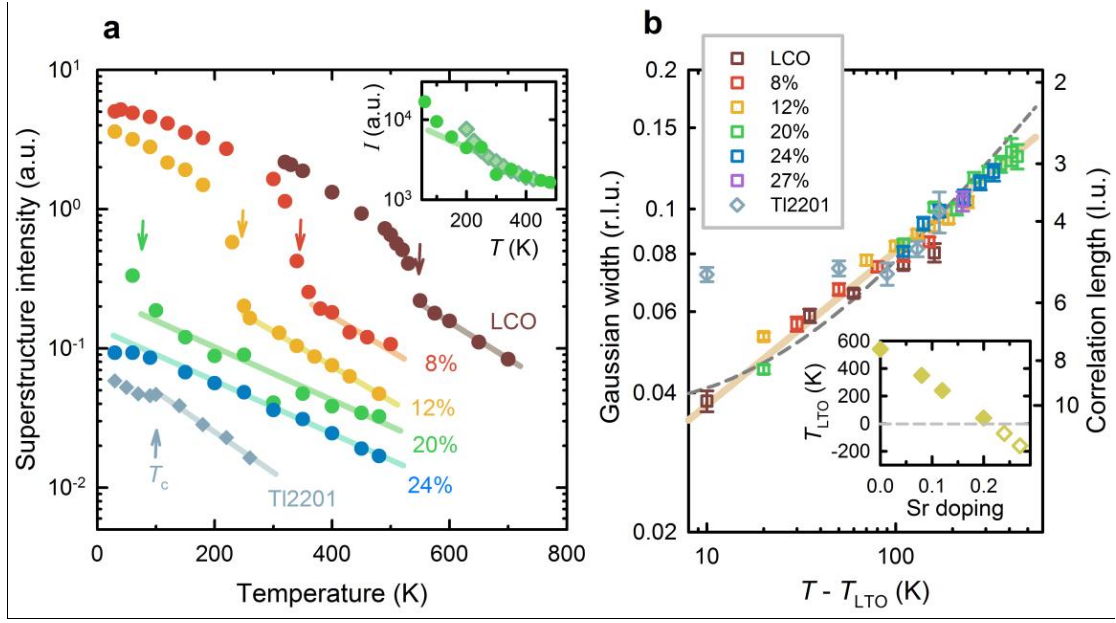

**Fig. 3: Universal behavior and scaling of short-range orthorhombic correlations**. **a** Temperature- and doping dependence of diffuse LTO superstructure intensity in $La_2CuO_4$ (LCO), $La_{2-x}Sr_xCuO_4$ (LSCO) and $Tl_2Ba_2CuO_{6+\delta}$ (Tl2201) obtained from X-ray scattering (see Methods for reciprocal space positions). Above the HTT-LTO transition at $T_{LTO}$ (arrows), the intensity decreases exponentially, at a nearly doping- and compound-independent rate of ~1/200 $K^{-1}$. The $x = 0.24$ LSCO sample does not show long-range LTO order, yet short-range correlations are still observable at 480 K. The Tl2201 sample is slightly overdoped, also shows no long-range orthorhombicity, and exhibits a clear change of the diffuse intensity below $T_c = 89$ K. Intensities for LSCO are normalized to the nearest Bragg peaks at 30 K, and therefore directly comparable. The Tl2201 data are shifted vertically by an arbitray constant for clarity. Inset: comparison between X-ray (circles, $x = 0.2$) and quasielastic neutron (diamonds, $x = 0.155$) diffuse intensity above $T_{LTO}$ reveals similar exponential behavior. The neutron data is for the (1.5 1.5 2) LTO peak. **b** Gaussian width of the diffuse LTO superstructure peaks *vs.* $T - T_{LTO}$. Master curve (solid line): $(T - T_{LTO})^{1/3}$; dashed line: shifted square-root dependence, $(T - T_{LTO}')^{1/2}$, as expected from rare-region theory, with $T_{LTO}' = T_{LTO} - 25$ K. The right axis shows the corresponding correlation length, in units of the in-plane tetragonal lattice constant. Inset: $T_{LTO}$ values used in the scaling, including the extrapolated effective negative values for the $x = 0.24$ and $0.27$ samples (empty diamonds). Interestingly, the Tl2201 data approximately follow the universal curve if the effective $T_{LTO}$ is taken to be ~$T_c$; the width remains roughly temperature-independent near and below $T_c$.

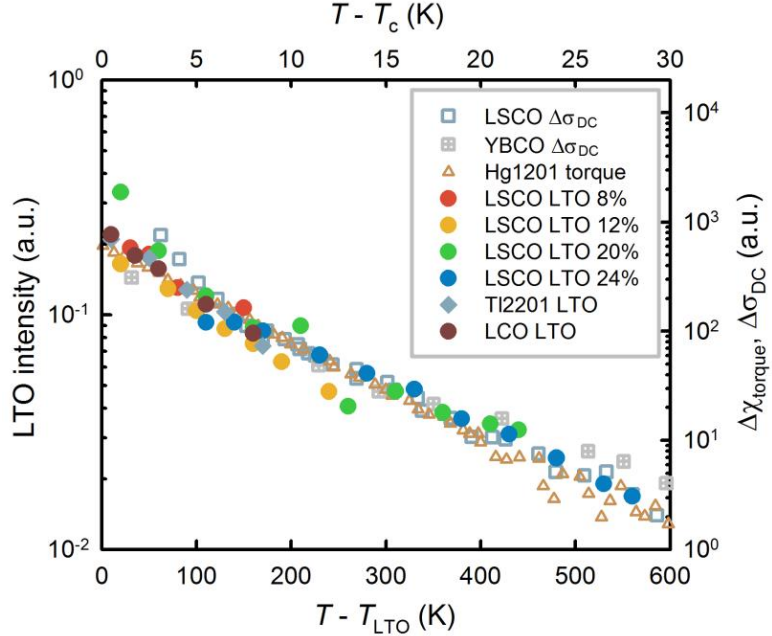

**Fig. 4: Comparison between orthorhombic and superconducting fluctuations.** The diffuse peak intensities normalized to the nearest Bragg peaks from Fig. 3a (left vertical scale, bottom horizontal scale) fall on the same universal exponential curve when the temperature is shifted by $T_{LTO}$, similar to the effective linewidths (Fig. 3b). The Tl2201 data have been multiplied by a constant to match the LSCO scaling. Signatures of superconducting fluctuations (right vertical scale, top horizontal scale) follow a similar exponential behavior when shifted by $T_c$, independent of cuprate family and doping. As representative examples, we show the superconducting contribution to the normal-state conductivity (paraconductivity, $\Delta\sigma_{DC}$) for LSCO with $x = 0.1$ ($T_c = 28$ K) [76] and $YBa_2Cu_3O_{6+\delta}$ (YBCO, $T_c = 93$ K) [77], as well as torque magnetometry results[15] for $HgBa_2CuO_{4+\delta}$ (Hg1201, $T_c = 70$ K). The same dependence is also observed using other probes, including specific heat and nonlinear conductivity/susceptibility (with different slopes for linear and nonlinear response)[12,13]. Neither YBCO nor Hg1201 exhibit LTO-type distortions, which implies that the structural and superconducting fluctuations are affected by a common, underlying cause. The scaling suggests that a hidden, universal inhomogeneity determines the behavior of both fluctuation regimes, leading to remarkable emergent simplicity.